\documentclass[copyright,creativecommons]{eptcs}
\providecommand{\event}{EXPRESS 2010}

\providecommand{\firstpage}{1}
\usepackage{breakurl}
\usepackage{times}
\usepackage{latexsym}
\usepackage{amsmath}
\usepackage{amssymb}
\usepackage{stmaryrd}
\usepackage{graphicx}

\title{A criterion for separating process calculi}

\author{Federico Banti \qquad\quad
Rosario Pugliese \qquad\quad
Francesco Tiezzi
\institute{Dipartimento di Sistemi e Informatica, Universit\`a degli Studi di Firenze,
Italy}
\email{fbanti@gmail.com\qquad rosario.pugliese@unifi.it\qquad tiezzi@dsi.unifi.it}
}

\def\titlerunning{A criterion for separating process calculi}
\def\authorrunning{F. Banti, R. Pugliese \& F. Tiezzi}

\newtheorem{example}{Example}[section]
\newtheorem{definition}{Definition}[section]
\newtheorem{lemma}{Lemma}[section]
\newtheorem{theorem}{Theorem}[section]

\newtheorem{proposition}{Proposition}[section]

\newcommand{\qed}{\ensuremath{\Box}}

\newenvironment{proofSketch}%
{\vspace{-.1cm}
{\it Proof (sketch).}
}%
{\qed
\smallskip
}

\newenvironment{Proof}%
{\vspace{-.1cm}
\noindent
{\it Proof.}
}%
{\qed
\smallskip
}

{\begin{trivlist}\refstepcounter{definition}%
\item[]{\bf Notation \thedefinition}}%
{\end{trivlist}}

\newcommand{\sep}{\quad | \quad} % Symbol used in grammar definitions to separate alternative forms
\newcommand{\pic}{\ensuremath{\pi}-calculus} % pi-calculus
\newcommand{\ccsg}{\ccs$^{sg}$}

\newcommand{\cpg}{CPG}
\newcommand{\ccsprio}{\ccs$^{\textsf{prio}}$}
\newcommand{\bccsp}{BCCSP$_{\Theta}$}

\newcommand{\ccs}{CCS}

\newcommand{\indhole}[1]{{\_\!\_}_{{}_{#1}}}

\makeatletter
\def \rightarrowfill{\m@th\mathord{\smash-}\mkern-6mu%
  \cleaders\hbox{$\mkern-2mu\mathord{\smash-}\mkern-2mu$}\hfill
  \mkern-6mu\mathord\rightarrow}
\makeatother
\newcommand{\transition}[1]{\overstackrel{\rightarrowfill}{\ \ #1\ \ \phantom{\taukill}}} % labelled transition
\newcommand{\ktransition}[2]{\overstackrel{\rightarrowfill}{\ \ #1\ \ \phantom{\taukill}}_{#2}}
\makeatletter
\def \wrightarrowfill{\m@th\mathord{\smash=}\mkern-6mu%
  \cleaders\hbox{$\mkern-2mu\mathord{\smash=}\mkern-2mu$}\hfill
  \mkern-6mu\mathord\Rightarrow}
\makeatother
\newcommand{\weakTransition}[1]{\overstackrel{\wrightarrowfill}{\ \ #1\ \ \phantom{\taukill}}} % weak labelled transition
\newcommand{\weakTransitionhat}[1]{\overstackrel{\wrightarrowfill}{\ \ \widehat{#1}\ \ \phantom{\taukill}}} % weak labelled transition
\newcommand{\kweakTransition}[2]{\overstackrel{\wrightarrowfill}{\ \ #1\ \ \phantom{\taukill}}_{#2}}
\makeatletter
\def \nrightarrowfill{\m@th\mathord{\smash-}\mkern-6mu%
  \cleaders\hbox{$\mkern-2mu\mathord{\smash-}\mkern-2mu$}\hfill
  \mkern-6mu\mathord\nrightarrow}
\makeatother
\def \overstackrel#1#2{\mathrel{\mathop{#1}\limits^{#2}}}





\newcommand{\wperform}[1]{\ensuremath{ \Downarrow_{#1} } }
\newcommand{\ghost} {\ensuremath{\not \Downarrow}}

\newcommand {\simuln} [1]  {\ensuremath{\preccurlyeq}^{#1}   }

\newcommand {\osimul}  {\ensuremath{ \preccurlyeq^\omega}   }

\newcommand{\tr} [1] {\ensuremath{[\![#1]\!]}}
\newcommand{\ie}{i.e.~}
\newcommand{\eg}{e.g.~}

\newcommand{\outDS}[1]{\langle #1 \rangle}

\newcommand{\outccs}[1]{ \overline{#1} }   % output prefix without object
\newcommand{\ren}[1]{ [ #1 ] }   % CCS renaming

\newcommand{\outpic}[2]{\overline{#1}\,#2}          %pic output
\newcommand{\inpic}[2]{#1(#2)}                      %pic input
\newcommand{\mat}[2]{[#1=#2]}                       %match
\newcommand{\reppic}{!}                             %replication
\newcommand{\fn}[1]{{\tt fn}(#1)}

\newcommand{\hood} [1] {\ensuremath{\ulcorner #1 \urcorner}}
\newcommand{\mpm}{\ensuremath{\pi^{{}_{MPM}}}-calculus}

\newcommand{\seq}[1]{\textbf{#1}}

\newcommand{\cows}{\textsc{COWS}} % COWS

\newcommand{\killing}[1]{\mathbf{kill}(#1)} % kill activity
\newcommand{\scope}[1]{[#1]\,} % delimitation
\newcommand{\nil}{\mathbf{0}} % nil
\newcommand{\arr}[1]{\langle #1 \rangle} % notation for array/tuple
\newcommand{\set}[1]{\{#1\}} % brackets for sets
\newcommand{\taukill}{\dag} % executed kill activity transtion label

\newdimen\proofrulebreadth \proofrulebreadth=.04em
\newdimen\proofdotseparation \proofdotseparation=1.25ex
\newdimen\proofrulebaseline \proofrulebaseline=2ex
\newcount\proofdotnumber \proofdotnumber=3
\let\then\relax
\def\hfi{\hskip0pt plus.0001fil}
\mathchardef\squigto="3A3B
\newif\ifinsideprooftree\insideprooftreefalse
\newif\ifonleftofproofrule\onleftofproofrulefalse
\newif\ifproofdots\proofdotsfalse
\newif\ifdoubleproof\doubleprooffalse
\let\wereinproofbit\relax
\newdimen\shortenproofleft
\newdimen\shortenproofright
\newdimen\proofbelowshift
\newbox\proofabove
\newbox\proofbelow
\newbox\proofrulename
\def\shiftproofbelow{\let\next\relax\afterassignment\setshiftproofbelow\dimen0 }
\def\shiftproofbelowneg{\def\next{\multiply\dimen0 by-1 }%
\afterassignment\setshiftproofbelow\dimen0 }
\def\setshiftproofbelow{\next\proofbelowshift=\dimen0 }
\def\setproofrulebreadth{\proofrulebreadth}

\def\prooftree{% NESTED ZERO (\ifonleftofproofrule)
\ifnum  \lastpenalty=1 \then   \unpenalty \else
\onleftofproofrulefalse \fi
\ifonleftofproofrule \else   \ifinsideprooftree
        \then   \hskip.5em plus1fil
        \fi
\fi
\bgroup% NESTED ONE (\proofbelow, \proofrulename, \proofabove,
\setbox\proofbelow=\hbox{}\setbox\proofrulename=\hbox{}%
\let\justifies\proofover\let\leadsto\proofoverdots\let\Justifies\proofoverdbl
\let\using\proofusing\let\[\prooftree
\ifinsideprooftree\let\]\endprooftree\fi
\proofdotsfalse\doubleprooffalse
\let\thickness\setproofrulebreadth
\let\shiftright\shiftproofbelow \let\shift\shiftproofbelow
\let\shiftleft\shiftproofbelowneg
\let\ifwasinsideprooftree\ifinsideprooftree
\insideprooftreetrue
\setbox\proofabove=\hbox\bgroup$\displaystyle % NESTED TWO
\let\wereinproofbit\prooftree
\shortenproofleft=0pt \shortenproofright=0pt \proofbelowshift=0pt
\onleftofproofruletrue\penalty1 }

\def\eproofbit{% NESTED TWO
\ifx    \wereinproofbit\prooftree \then   \ifcase \lastpenalty
        \then   \shortenproofright=0pt  % 0: some other object, no indentation
        \or     \unpenalty\hfil         % 1: empty hypotheses, just glue
        \or     \unpenalty\unskip       % 2: just had a tree, remove glue
        \else   \shortenproofright=0pt  % eh?
        \fi
\fi
\global\dimen0=\shortenproofleft \global\dimen1=\shortenproofright
\global\dimen2=\proofrulebreadth \global\dimen3=\proofbelowshift
\global\dimen4=\proofdotseparation
$\egroup  % NESTED ONE
\shortenproofleft=\dimen0 \shortenproofright=\dimen1
\proofrulebreadth=\dimen2 \proofbelowshift=\dimen3
\proofdotseparation=\dimen4
}

\def\proofover{% NESTED TWO
\eproofbit % NESTED ONE
\setbox\proofbelow=\hbox\bgroup % NESTED TWO
\let\wereinproofbit\proofover
$\displaystyle
}%
\def\proofoverdbl{% NESTED TWO
\eproofbit % NESTED ONE
\doubleprooftrue
\setbox\proofbelow=\hbox\bgroup % NESTED TWO
\let\wereinproofbit\proofoverdbl
$\displaystyle
}%
\def\proofoverdots{% NESTED TWO
\eproofbit % NESTED ONE
\proofdotstrue
\setbox\proofbelow=\hbox\bgroup % NESTED TWO
\let\wereinproofbit\proofoverdots
$\displaystyle
}%
\def\proofusing{% NESTED TWO
\eproofbit % NESTED ONE
\setbox\proofrulename=\hbox\bgroup % NESTED TWO
\let\wereinproofbit\proofusing
\kern0.3em$ }

\def\endprooftree{% NESTED TWO
\eproofbit % NESTED ONE
  \dimen5 =0pt% spread of hypotheses
\dimen0=\wd\proofabove \advance\dimen0-\shortenproofleft
\advance\dimen0-\shortenproofright
\dimen1=.5\dimen0 \advance\dimen1-.5\wd\proofbelow \dimen4=\dimen1
\advance\dimen1\proofbelowshift \advance\dimen4-\proofbelowshift
\ifdim  \dimen1<0pt \then   \advance\shortenproofleft\dimen1
        \advance\dimen0-\dimen1
        \dimen1=0pt
        \ifdim  \shortenproofleft<0pt
        \then   \setbox\proofabove=\hbox{%
                        \kern-\shortenproofleft\unhbox\proofabove}%
                \shortenproofleft=0pt
        \fi
\fi
\ifdim  \dimen4<0pt \then   \advance\shortenproofright\dimen4
        \advance\dimen0-\dimen4
        \dimen4=0pt
\fi
\ifdim  \shortenproofright<\wd\proofrulename \then
\shortenproofright=\wd\proofrulename \fi
\dimen2=\shortenproofleft \advance\dimen2 by\dimen1
\dimen3=\shortenproofright\advance\dimen3 by\dimen4
\ifproofdots \then
        \dimen6=\shortenproofleft \advance\dimen6 .5\dimen0
        \setbox1=\vbox to\proofdotseparation{\vss\hbox{$\cdot$}\vss}
        \setbox0=\hbox{%
                \kern\dimen6
                $\vcenter to\proofdotnumber\proofdotseparation
                        {\leaders\box1\vfill}$%
                \unhbox\proofrulename}%
\else   \dimen6=\fontdimen22\the\textfont2 % height of maths axis
        \dimen7=\dimen6
        \advance\dimen6by.5\proofrulebreadth
        \advance\dimen7by-.5\proofrulebreadth
        \setbox0=\hbox{%
                \kern\shortenproofleft
                \ifdoubleproof
                \then   \hbox to\dimen0{%
                        $\mathsurround0pt\mathord=\mkern-6mu%
                        \cleaders\hbox{$\mkern-2mu=\mkern-2mu$}\hfill
                        \mkern-6mu\mathord=$}%
                \else   \vrule height\dimen6 depth-\dimen7 width\dimen0
                \fi
                \unhbox\proofrulename}%
        \ht0=\dimen6 \dp0=-\dimen7
\fi
\let\doll\relax
\ifwasinsideprooftree \then   \let\VBOX\vbox \else
\ifmmode\else$\let\doll=$\fi
        \let\VBOX\vcenter
\fi
\VBOX   {\baselineskip\proofrulebaseline \lineskip.2ex
        \expandafter\lineskiplimit\ifproofdots0ex\else-0.6ex\fi
        \hbox   spread\dimen5   {\hfi\unhbox\proofabove\hfi}%
        \hbox{\box0}%
        \hbox   {\kern\dimen2 \box\proofbelow}}\doll%
\global\dimen2=\dimen2 \global\dimen3=\dimen3
\egroup % NESTED ZERO
\ifonleftofproofrule \then   \shortenproofleft=\dimen2 \fi
\shortenproofright=\dimen3
\onleftofproofrulefalse \ifinsideprooftree \then   \hskip.5em plus
1fil \penalty2 \fi }

\begin{document}

\maketitle

\begin{abstract}
We introduce a new criterion, replacement freeness, to discern the
relative expressiveness of process calculi. Intuitively, a calculus
is strongly replacement free if replacing, within an enclosing
context, a process that cannot perform any visible action by an
arbitrary process never inhibits the capability of the resulting
process to perform a visible action. We prove
that there exists no compositional and interaction sensitive
encoding of a not strongly replacement free calculus into any
strongly replacement free one. We then define a weaker version of
replacement freeness, by only considering replacement of closed
processes, and prove that, if we additionally require the
encoding to preserve name independence, it is not even possible to
encode a non replacement free calculus into a weakly replacement
free one. As a consequence of our encodability results, we
get that many calculi equipped with priority are not
replacement free and hence are not encodable into mainstream calculi
like \ccs\ and \pic, that instead are strongly replacement free. We
also prove that variants of \pic\ with match among names, pattern
matching or polyadic synchronization are only weakly replacement
free,
hence they are separated both from process calculi with
priority and from mainstream calculi.
\end{abstract}

\setcounter{footnote}{0}

\section{Introduction}
\label{sec:introduction}

The field of process calculi has been sometimes compared to a
`jungle' of interrelated but separate
theories~\cite{Nestmann06,parrow-expressive08},
made of plenty of calculi, each one with its own set of concepts,
operators, semantics and results. With the aim of turning this
jungle into a `nicely organized garden', many authors have tackled
the challenge of devising suitable criteria to classify the
different calculi. Relative expressiveness has been then advocated
as a valid perspective from which two calculi can be compared.
A standard approach is to define a `proper' encoding of a calculus
into another one, \ie a function mapping terms of the source
calculus into terms of the target one that is required to
preserve and/or reflect `reasonably' much of the semantics of
the source language and to be structurally defined over
its operators. The target calculus is then considered at least
as expressive as the source one. Alternatively, one can prove
a sort of \emph{separation} result stating that no such
encoding exists, thus telling the two calculi apart.

This is an effective approach, but there is no common
agreement on which class of encodings has to be used.
Several different classes have been introduced (see, e.g.,
\cite{priority1999,Carbone-express-polyadic03,Palamidessi-expressive-synch03,Gorla08:encodability-separation,Gorla-reasonable-main08,parrow-expressive08,priority-guards08,Aranda-Versari-expressivity09,Nadia-memorial-expr09,FuLu:10}),
each one being characterised by the syntactic and semantic
properties that the encodings are required to satisfy.
Appropriateness of a class depends however from
the kind of results one is seeking. Encodings are better, in the
sense that they attest that the target calculus has expressive
power tighter to that of the source calculus, when satisfying
as many properties as possible. Conversely, separation results are
stronger and more informative when relying on encodings with minimal
requirements.

In this paper we introduce a few criteria and classes of encodings for
separating process calculi, and illustrate some results of their
application. In particular, we separate extensions of \pic\ from the
core calculus and calculi with priority mechanisms from the others.

Since our focus is on separating process calculi, to get more general
results we rely on a minimal set of requirements taken from
the literature.
The starting point of our investigation are the \emph{reasonable
encodings}, introduced in~\cite{Gorla-reasonable-main08}
for comparing several communication primitives in the context of \pic,
We further generalise this already broad class of encodings by
dropping the requirements about name invariance, operational
correspondence, and divergence preservation and reflection.
Thus, we get the class of \emph{basic encodings}, \ie encodings
that only require the two basic properties that, in our opinion,
any encoding should satisfy:
compositionality (\ie the encoding of a compound term is defined
by combining the encodings of its sub-terms) and
interaction sensitiveness (\ie the capability to interact with
the context through visible actions is preserved and reflected).

We first introduce a new criterion for separating process calculi, named
\emph{replacement freeness}.
Intuitively, a calculus is strongly replacement free
(\emph{strongly rep-free}, for short) if replacing,
within an enclosing context, an `invisible' process
(\ie a process that cannot perform any visible action) by an
arbitrary process never inhibits the capability of the resulting
process to perform a visible action.
We then prove that there exist no basic encodings of non strongly
rep-free calculi into strongly rep-free ones.
Of course, a similar result also holds for reasonable
encodings, since they are basic encodings too.

\begin{figure}[t]
\centering
\includegraphics[scale=.46]{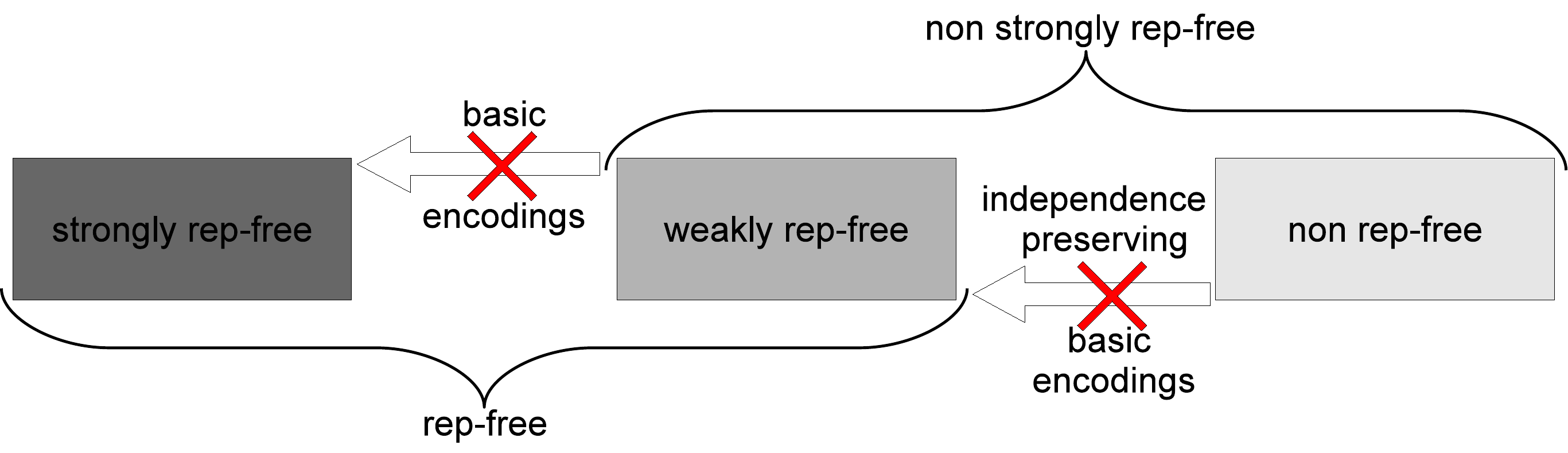}
\vspace*{-.5cm}
\label{fig:partition}
\caption{Calculi partition}
\vspace*{-.5cm}
\end{figure}

Intuitively, invisible processes cannot explicitly interact with the
enclosing context since they do not perform visible actions (at most,
they can only perform `internal' computation steps). Nevertheless,
their behaviour could be implicitly affected by the enclosing context
through the generation of substitutions involving the free names of the
invisible processes. To prevent also this kind of influence, we will
consider the subclass of invisible processes that are closed (\ie contain
no free name) and use it to weaken the condition of replacement freeness.
A calculus is hence deemed replacement free (\emph{rep-free},
for short) if replacing, within an enclosing context, a closed
invisible process by an arbitrary process never inhibits
the capability of the resulting process to perform visible actions.
Of course any strongly rep-free calculus is also rep-free;
we will show that the converse does not hold. We will call
\emph{weakly rep-free} those rep-free calculi that are not
strongly rep-free. Since the processes which we now focus on
share no free names, we limit ourselves to only consider the
subclass of basic encodings that preserve
\emph{name independence} \cite{Palamidessi-expressive-synch03,PhillipsV06}
(\ie if two processes share no free names the same holds for their encodings).
We will prove that there exist no such encodings of non rep-free
calculi into (even weakly) rep-free calculi. In the end, we obtain a
de facto tripartition of process calculi into three sets,
i.e. strongly rep-free, weakly rep-free, and non rep-free calculi,
which are respectively separated by basic encodings and
independence preserving basic encodings
(as shown in Figure~\ref{fig:partition}).

Then, we will present several results, arising from the exploitation
of our criteria, about the relative expressiveness of well-known
process calculi. We first prove that some mainstream
process calculi, like \ccs~\cite{CCS} and \pic~\cite{PICALC}, are strongly rep-free.
Conversely, most of the calculi with some form of priority that
have been proposed in the literature (\eg those
in~\cite{aceto-priority08,priority-guards08,priority1999,LPT07:ESOP}),
are not rep-free and thus cannot be `properly' encoded into \ccs\ or \pic.
Intuitively, when replacing an invisible process with one performing
prioritized actions, replacement freeness can be violated because
the additional initial actions at disposal of the replacing process
may prevent some actions with lower priority that might originally
be performed.
However, this turns out not to be the only source of possible
violations of replacement freeness. Indeed, we also show that variants
of \pic\ equipped with, respectively, match among names~\cite{PICALC},
polyadic synchronization~\cite{Carbone-express-polyadic03} or
pattern matching~\cite{Gorla-reasonable-main08} are weakly rep-free
and thus are strictly more expressive than the `classical' \pic.
This allows us to prove in a quite simple and uniform way
possibly stronger versions of the results obtained in
\cite{priority-guards08,Carbone-express-polyadic03,Gorla08:encodability-separation,Gorla-reasonable-main08,Nadia-memorial-expr09}. As concerns these calculi, strong re\-place\-ment freeness is violated
because a process originally invisible can be transformed into a visible
one via application of a name substitution (generated by the enclosing
context) which originates a new computation.
The reason for this class of violations is clearly different from the
previously discussed one and, in fact, if only closed invisible processes
are taken into account, these violations do not arise. Thus, the above
mentioned richer variants of \pic\ are only weakly rep-free.
Anyway, they cannot encode non rep-free calculi like those with
priority herein analyzed.

The rest of the paper is structured as follows.
Section~\ref{sec:criterion} introduces the replacement freeness
criterion, both in its stronger and in its weaker formulation, and
the classes of basic and independence preserving encodings
we exploit;
it also presents our general separation results.
Section~\ref{sec:rep-free} proves that \ccs\ and \pic\ are strongly rep-free,
Section~\ref{sec:weakly-repfree} proves that some variants of \pic\
are only weakly rep-free, and Section~\ref{sec:non-repfree} proves
that several calculi with priority are not (even weakly) rep-free.
Finally, Section~\ref{sec:conclusion}
draws a few conclusions and
reviews some strictly related work.
We refer the interested reader to \cite{ExprFULL} for
a full account of the proofs and for a wider comparison
with related work.

\section{Replacement freeness: a separation criterion}
\label{sec:criterion}

Conceptually speaking, our approach relies on some properties
of process calculi that are
invariant under certain classes of encodings, so that, if an encoding
violates one such invariant, it cannot belong to the intended class.
Strong replacement freeness, or, to be precise, its negation,
is an invariant for basic encodings, as well as negation of
replacement freeness
is an invariant for independence preserving basic encodings.
From the quite simple concepts and results presented in this section
it follows a powerful methodology for separating process calculi:
non strongly rep-free calculi cannot be encoded through basic
encodings into strongly rep-free calculi; furthermore,
%there are no independence preserving basic encodings
%of non rep-free calculi into (possibly weakly) rep-free ones.
non rep-free calculi cannot be encoded through independence
preserving basic encodings into (possibly weakly) rep-free calculi.
In the next sections, we present several results about
the relative expressiveness of well-known process calculi
arising from the exploitation of our methodology.

\subsection{Background notations}
\label{sec:background}

Process calculi are formal languages allowing to constructs
operational models of open computing systems and to specify
interactions between systems. They provide different sets of
operators for composing terms, called \emph{processes}, as well as
different sets of \emph{(atomic) actions}, typically representing
inputs and outputs along communication channels, that processes can
perform.

Actions are ranged over by $\mu,\mu',\mu_1,\ldots$ and may be either
\emph{visible} (we use $\alpha,\alpha',\beta,\ldots$ to range over
them) or \emph{invisible} (in which case they are indistinguishable
and usually denoted only by $\tau$).
Specifically, actions are expressed in terms of \emph{names}, \ie
basic entities without structure, ranged over by letters $a$, $b$,
\ldots, $x$, $y$, \ldots, $n$, $m$, \ldots.
To define and delimit the scope of names, process calculi are
equipped with \emph{name-binding} operators.
An occurrence of a name in a process is \emph{bound} if it is, or it
lies within the scope of, a binding occurrence of the name. An
occurrence of a name in a process is \emph{free} if it is not bound.
We write $\fn{P}$ for the set of names that have a free occurrence in $P$.
A process $P$ is \emph{closed} if $\fn{P}=\emptyset$.

For any given process calculus, we need a notion of \emph{context}
where a term of the calculus can be placed for execution. Although
we will usually deal with contexts with a single hole, we need to
introduce the more general notion of $k$-hole context.
\begin{definition}[$k$-hole context]
\label{def:kholecontext} A \emph{$k$-hole context}, with $k \geq 1$,
is a term of the calculus where $k$ sub-terms are replaced by the
holes $\indhole{1}$, \ldots, $\indhole{k}$. If $C$ is a $k$-hole
context then we write $C[P_1,\ldots,P_k]$ for the term obtained by
replacing $\indhole{i}$ in $C$ by $P_i$, for $i \in [1..k]$.
We also write $\indhole{}$ in place of $\indhole{1}$ for $1$-hole contexts.
\end{definition}

We shall use $\mathbb{C}$, $\mathbb{C}_1$, $\mathbb{C}_2$,
\ldots to range over process calculi. When convenient,
we shall regard a calculus as a set of processes, contexts and operators,
writing \eg $P\in\mathbb{C}$ to mean that $P$ is an element
of $\mathbb{C}$.

We assume that the operational semantics of process calculi
is defined by means of \emph{labelled transition systems}.
Transitions labelled by invisible actions correspond to
\emph{computation steps} and can be thought of as taking place of
`internal' interactions of systems, whereas transitions labelled by
visible actions can be thought of as representing only `potential'
computation steps, since in order for them to occur they require
a contribution from the environment.
As usual, we will write $P \transition{\mu} P'$ to indicate
that the process $P$ can do a transition labelled $\mu$ and
become the process $P'$ in doing so.
We let $\weakTransition{}$ to denote the reflexive and transitive
closure of $\transition\tau$, $\weakTransition\mu$ to denote
$\weakTransition{} \transition\mu \weakTransition{}$
(the juxtaposition of two relations indicates their composition),
and $\weakTransitionhat{\mu}$ to denote $\weakTransition{}$,
if $\mu = \tau$, and $\weakTransition\mu$, otherwise.
Moreover, we will write $\ktransition{\mu}{k}$ to denote the
composition of $\transition{\mu}$ with itself $k$-times
(similarly for the other transition relations).

Now, by exploiting the relations above, we define the following predicates over processes.

\begin{definition}[Process predicates]
\label{processpredicates} Let $P$ be a process.
\begin{itemize}\vspace*{-.2cm}
\item $P \wperform\alpha$, \ie $P$ \emph{can perform the
    (visible) action} $\alpha$, if $P \weakTransition\alpha P'$
    for some $P'$;
\item $P \Downarrow$, \ie $P$ is \emph{visible}, if there exists a
    visible action $\alpha$ such that $P \wperform\alpha$;
\item $P \ghost$, \ie $P$ is \emph{invisible}, if there
    exists no visible action $\alpha$ such that $P
    \wperform\alpha$.
\end{itemize}
\end{definition}
Predicate $\wperform\alpha$ accounts for the ability of processes
of interacting with their environment.
Notably, an invisible process either is stuck or can only perform
invisible actions. Clearly, if $P$ is invisible, then it can
only evolve to invisible processes.

Our results also hold for process calculi whose operational
semantics is defined by means of \emph{reduction relations}.
In this case, predicate $\wperform\alpha$ is defined by
induction on the syntax of processes.

\subsection{Basic encodings and strong replacement freeness}
\label{sec:basicenc}

Basic encodings take their appellation from the basic
properties they are required to satisfy. The first requirement
regards the `structure' of the source calculus: an encoding must be
\emph{compositional}, \ie every $k$-ary operator $\mathit{op}$ of
the source calculus is translated into a $k$-hole context $C$ of the
target calculus and the application of $\mathit{op}$ to $k$
processes is encoded into the application of $C$ to the encodings of
such processes. The second requirement regards the `semantics' of
the source calculus: an encoding must be \emph{interaction
sensitive}, \ie it must preserve and reflect the capability of a
process to perform, or not, visible actions (possibly after some
internal computation steps).

\begin{definition}[Basic encodings]
\label{def:encoding}
An encoding $\tr \cdot$ of $\mathbb{C}_1$ into $\mathbb{C}_2$ is
\emph{basic} if
\begin{itemize}
\item $\tr \cdot$ is \emph{compositional}: for every $k$-ary operator
    $op\ \in \mathbb{C}_1$ there is a $k$-hole context $C_{op}$
    $\in \mathbb{C}_2$ such that
    $\forall\ P_1, \ldots,\ P_k \ \in\ \mathbb{C}_1$,
    $\tr{op(P_1,\ldots,P_k)} = C_{op}[\tr{P_1},\ldots,\tr{P_k}]\,.$
\item $\tr \cdot$ is \emph{interaction sensitive}: for every process
    $P\ \in \mathbb{C}_1$, $P \Downarrow$ if and only if $\tr P
    \Downarrow\,$.
\end{itemize}
\end{definition}

Notice that the property of being an invisible process is an
invariant under basic encodings, since by definition such
encodings preserve and reflect processes interaction ability.

\begin{proposition}
\label{prop:ghost}
Let $\tr \cdot$ be a basic encoding of $\mathbb{C}_1$ into $\mathbb{C}_2$
and $P\in\mathbb{C}_1$.
Then $P\ghost\;$ if, and only if, $\tr{P}\ghost\:$.
\end{proposition}

By using a basic encoding (in fact, a compositional
one would suffice), one can encode not only single
operators but also arbitrary contexts. In other words, any
context in $\mathbb{C}_1$ can be represented as a context in
$\mathbb{C}_2$. We state this property for 1-hole
contexts only, but it could be easily generalised.

\begin{lemma}
\label{lemma:context-encoding}
Let $\tr \cdot$ be a basic encoding of $\mathbb{C}_1$ into $\mathbb{C}_2$.
Then, for every context $C_1\in\mathbb{C}_1$, there exists a context
$C_2\in\mathbb{C}_2$ such that, for every process $P\in\mathbb{C}_1$,
$\tr{C_1[P]}  =  C_2[\tr P] .$
\end{lemma}
\begin{proofSketch}
By induction on the structure of $C_1$.
\end{proofSketch}

A strongly rep-free calculus is a calculus for which, the replacement
within a context of an invisible process with any other one, never
inhibits the capability of executing a visible action.

\begin{definition}[Strong replacement freeness]
\label{def:rep-free}
A calculus $\mathbb{C}$
is \emph{strongly replacement free} (\emph{strongly rep-free}, for
short) if for every context $C$, invisible process $I$ and process
$P$ in $\mathbb{C}$,
\vspace*{-.2cm}
\begin{eqnarray}
\label{cond:rep-free} C[I]\Downarrow\quad & \mbox{implies} & \quad
C[P]\Downarrow
\end{eqnarray}
\end{definition}

A calculus is not strongly rep-free if there exists a triple $C$,
$I$ and $P$ violating the condition (\ref{cond:rep-free}).

\begin{example}
\label{ex:picMatch}
Consider \pic\ with the \emph{match} operator $\mat{x}{y} P$,
which allows a process to test if the names $x$ and $y$ coincide and to continue its execution as $P$ only if the test succeeds
(see also Section~\ref{sec:weakly-repfree}).
This calculus is not strongly rep-free, because of the
following triple:
$C = (\nu x)(\inpic{x}{a} . \indhole{} \mid \outpic{x}{b})$,
$I = \mat{a}{b} \outpic{y}{c}$ and $P = \nil$.
Process $I$ is invisible since it is blocked by an unsatisfied
match (as names $a$ and $b$ are supposed to be different). However,
we have that $C[I] \Downarrow$, since
$
C[I] = (\nu x)(\inpic{x}{a} . \mat{a}{b} \outpic{y}{c} \mid
\outpic{x}{b}) \transition \tau (\nu x)(\mat{b}{b} \outpic{y}{c}
\mid \nil) \transition{\outpic{y}{c}} (\nu x)(\nil \mid \nil)
$.
Instead, $C[P] \ghost$ because $C[P]$ can only perform the transition
$
C[P] = (\nu x)(\inpic{x}{a} \mid \outpic{x}{b}) \transition \tau
(\nu x)(\nil \mid \nil)
$
and become stuck in doing so.
\end{example}

The proof of the following separation result is based on showing
that the property of being a non strongly rep-free calculus is
invariant under basic encodings.

\begin{theorem}
\label{theo:rep-free}
There is no basic encoding from a non strongly rep-free to a strongly rep-free calculus.
\end{theorem}
\begin{Proof}
Let $\mathbb{C}_1$ and $\mathbb{C}_2$ be two process calculi,
let $\mathbb{C}_2$ be strongly rep-free and $\mathbb{C}_1$ be not.
Let us assume that $\tr \cdot$ is a
basic encoding of $\mathbb{C}_1$ into $\mathbb{C}_2$. Let $C$,
$I$ and $P$ in $\mathbb{C}_1$ be such that $C[I]\Downarrow$
and $C[P]\ghost$; such a triple exists since $\mathbb{C}_1$ is not
strongly rep-free. By Proposition~\ref{prop:ghost}, $\tr I$ is
invisible.
By Lemma~\ref{lemma:context-encoding}, for some context $C'$ of
$\mathbb{C}_2$, we have $\tr{C[I]}=C'[\tr I]$ and $\tr{C[P]} = C'[\tr
P]$. Thus, by interaction sensitiveness and Proposition~\ref{prop:ghost},
$C'[\tr I]\Downarrow$ and $C'[\tr P]\ghost$ against the initial
assumption that $\mathbb{C}_2$ is strongly rep-free.
\end{Proof}

\subsection{Independence preserving basic encodings and replacement freeness}
\label{sec:independence}

The set of process calculi violating strong replacement freeness is
indeed quite large and can be further split into two distinct sets.
One set comprises calculi for which an invisible process $I$ may be
transformed into a visible one by application of a
\emph{substitution} $\sigma$ (i.e. a function on names).
As shown in Section~\ref{sec:weakly-repfree}, this happens for
calculi exploiting such operators as, e.g., match among names,
polyadic synchronization, or pattern matching. The other set
includes, at least, those calculi exploiting some form of
priority, as shown in Section~\ref{sec:non-repfree}.
It is possible to formally separate these two sets of calculi by
defining a weaker version of replacement freeness based on the
subset of invisible processes that are also closed, \ie without
free names, and hence impervious to substitutions.
Intuitively, there is no way for the enclosing context to affect
the behaviour of closed invisible processes.
We will thus show that
the variants of \pic\ presented in Section~\ref{sec:weakly-repfree}
turn out to be (weakly) rep-free, but not strongly rep-free.
Instead, the process calculi with priority presented in
Section~\ref{sec:non-repfree} are not (even weakly) rep-free.
To prove separation of these two sets of calculi we only consider
those basic encodings that also preserve name
independence~\cite{Palamidessi-expressive-synch03,PhillipsV06},
\ie guarantee that if two processes do not share free names, the
same holds for their encodings.

\begin{definition}[Name independence]
\label{def:independece}
Two processes $P$ and $Q$ are \emph{independent} if $\fn{P} \cap
    \fn{Q} = \emptyset$.
An encoding $\tr \cdot$ of $\mathbb{C}_1$ into $\mathbb{C}_2$
    is \emph{independence preserving} if
    whenever processes $P$ and $Q$ in $\mathbb{C}_1$ are independent,
    then $\tr P$ and $\tr Q$ in $\mathbb{C}_2$ are independent too.
\end{definition}

Since closed processes have no free names and substitutions only
apply to free names, we get that if $P$ is a closed process, then
$P\sigma = P$, for any substitution $\sigma$.
Similarly, if $I$ is a closed invisible process,
then $I\sigma$ is a closed invisible process too.
Moreover, we can show that the property of a process to be closed is
invariant under independence preserving encodings.

\begin{lemma}
\label{lemma:bounded}
Let $\tr \cdot$ be an independence preserving encoding of
$\mathbb{C}_1$ into $\mathbb{C}_2$ and let $P\in\mathbb{C}_1$
be a closed process. Then $\tr P \in\mathbb{C}_2$ is closed too.
\end{lemma}
\begin{Proof}
By contradiction, let $P$ be a closed process such that
$\tr P$ is not closed. Then, by definition of closed process
(Definition~\ref{def:independece}), we would get that $\fn{P} =
\emptyset$ and, hence, $\fn{P} \cap \fn{P} = \emptyset$. Similarly, we
would get that $\fn{\tr P} \neq \emptyset$ and, hence, $\fn{\tr P}
\cap \fn{\tr P} \neq \emptyset$. Therefore, we would obtain that
processes $P$ and $P$ would be independent while $\tr P$ and $\tr P$
would not be, against the hypothesis that the encoding $\tr \cdot$
is independence preserving.
\end{Proof}

The weak version of replacement freeness is defined by imposing
condition (\ref{cond:rep-free}) on triples $C$, $I$ and $P$, where $I$
is a closed invisible process.

\begin{definition}[Replacement freeness]
\label{def:weak-rep-free}
A calculus $\mathbb{C}$
is \emph{replacement free} (\emph{rep-free}, for short)
if for every context $C$, closed invisible process $I$
and process $P$ in $\mathbb{C}$,
\vspace*{-.25cm}
\begin{eqnarray}
\label{cond:weak-rep-free}
C[I] \Downarrow  \ \ \ implies \ \ \  C[P]\Downarrow
\end{eqnarray}
\vspace*{-.7cm}\\
A calculus that is rep-free, but not strongly rep-free
is called \emph{weakly rep-free}.
\end{definition}

\begin{example}
\label{ex:ccsg}
Consider \ccsg, an extension of \ccs\ where channels have
priority levels and only complementary actions at the same level of
priority can synchronise (see Section~\ref{sec:non-repfree}
for more details).
We consider here just two priority levels, ordinary actions
and higher priority, underlined actions, so that ordinary actions
are preempted by taking place of synchronization between
high-priority actions.
\ccsg\ is not rep-free because of the following triple:
$C = ({\underline a} \mid \indhole{}) \backslash \set{\underline a} + \outccs{b}$, $I = \nil$ and $P = \outccs{\underline a}$.
In fact, we have that $C[I] \Downarrow$, because
$
C[I] = ({\underline a} \mid \nil) \backslash \set{\underline a} + \outccs{b}
\transition{\outccs{b}}
\nil
$, while $C[P] \ghost$, since process $C[P]$ can only
perform the transition
$
C[P] = ({\underline a} \mid \outccs{\underline a}) \backslash
\set{\underline a} + \outccs{b} \transition{\underline\tau} (\nil
\mid \nil) \backslash \set{\underline a}
$
and become stuck in doing so.
\end{example}

Of course, every strongly rep-free calculus is also rep-free and
every non rep-free calculus is also not strongly rep-free
(see Figure~\ref{fig:partition}). Finally, it is possible to
obtain the analogous separation result of
Theorem~\ref{theo:rep-free} for rep-free calculi by further
requiring the basic encodings to be independence preserving.

\begin{theorem}
\label{theo:weak-rep-free}
There is no independence preserving basic encoding from a
non rep-free to a rep-free calculus.
\end{theorem}
\begin{Proof}
The proof proceeds like that of Theorem~\ref{theo:rep-free} but
exploiting the hypothesis that $I$ is closed and that the property
of being closed is
invariant under independence preserving
encodings (Lemma~\ref{lemma:bounded}).
\end{Proof}

\section{Proving strong replacement freeness of mainstream calculi}
\label{sec:rep-free}

In this section we prove that the two well-known process calculi
\ccs\ and \pic\ are strongly rep-free. We do this by defining a family of
relations $\simuln{k}$ for $k < \omega$ (where $\omega$ is the
first infinite ordinal) and using them for proving, by induction
on $k$, that, both for \ccs\ and \pic, it holds that
$C[I] \osimul C[P]$, for every context $C$, invisible process
$I$ and process $P$. Proposition~\ref{prop:simulate-rep-free}
will then allow us to conclude.

\begin{definition}[$\omega$-simulation]
\label{def:osimulation}\mbox{}\vspace*{-.1cm}
\begin{enumerate}
\item $\simuln 0$ is the universal relation on processes.
\\[-.5cm]

\item For $1 \leq k < \omega$,
$\simuln{k}$ is defined by:\\
    \qquad
    $Q \simuln k P$ if, whenever $Q\transition{\mu} Q'$, then, for some $P'$, $P\weakTransitionhat \mu  P'$ and $Q' \simuln{k-1} P'$.
\\[-.5cm]

\item Relation $Q \osimul P$ (also referred as \emph{$\omega$-simulation}) holds if $Q \simuln k P$ for every $k$.
    If $Q \osimul P$ then we say that $P$ is an \emph{$\omega$-simulation} of $Q$ or that $P$ \emph{$\omega$-simulates} $Q$.
\\[-.5cm]
\end{enumerate}
\end{definition}

Intuitively, if $P$ $\omega$-simulates $Q$, then $P$ can perform
\emph{at least} the same visible actions that $Q$ can perform,
possibly preceded and followed by invisible actions, and the same
holds for their derivatives.

\label{stratification}
The above relations resemble the `stratification' of
weak bisimulation for \pic\ \cite[page 99]{PICALCBOOK}. However,
our relations are not symmetric and are used as a viable
technique for proving that \ccs\ and \pic\ are strongly rep-free,
rather than to capture observational equivalences among processes.
Given the results in \cite{CCS,PICALCBOOK} about the stratification
of weak bisimulation, we can also presume that, in \ccs\ and \pic,
$\osimul$ coincides with the standard \emph{simulation preorder}.

The following proposition states that $C[I] \osimul C[P]$ implies
conditions (\ref{cond:rep-free}) of Definition~\ref{def:rep-free}
and (\ref{cond:weak-rep-free}) of Definition~\ref{def:weak-rep-free}.
It then provides a technique for proving (strongly) replacement freeness,
not an alternative characterization.
In fact, condition $C[I] \osimul C[P]$ is stronger than
that requested in the (strongly) rep-free definition, since the former
requires that if $C[I]$ performs a visible action then $C[P]$ must
perform the same action, while the latter only requires that
$C[P]$ is able to perform some visible action.

\begin{proposition}[$\omega$-simulations \& replacement freeness]
\label{prop:simulate-rep-free} Let $\mathbb{C}$ be a process
calculus.
\begin{enumerate}
\item If, for every context $C$, invisible process $I$ and process $P$,
    it holds that $C[I] \osimul C[P]$ then $\mathbb{C}$ is a
    strongly rep-free calculus.
\\[-.5cm]
\item If, for every context $C$, closed invisible process $I$ and
    process $P$, it holds that $C[I] \osimul C[P]$ then $\mathbb{C}$
    is a rep-free calculus.
\\[-.5cm]
\end{enumerate}
\end{proposition}
\begin{Proof}
We only prove the thesis for case \emph{1} as the other case is similar.
Let $P$, $I$ and $C$ be a triple such that $C[I] \osimul C[P]$ and
$C[I] \Downarrow$.
To prove that $\mathbb{C}$ is strongly rep-free, we must show that
$C[P] \Downarrow$. In fact, $C[I] \Downarrow$ means that
$C[I] \wperform \alpha$ for some visible action $\alpha$.
Hence, for some $m \geq 0$ and processes $Q'$ and $Q''$, $C[I]
\ktransition{\tau}{m} Q' \transition \alpha Q''$. Since $C[I]
\osimul C[P]$ implies, in particular, that $C[I] \simuln{m+1} C[P]$,
then, by applying $m+1$ times Definition~\ref{def:osimulation} we
get that $C[P] \kweakTransition{}{m} R' \weakTransition \alpha R''$
for some processes $R'$ and $R''$. Hence, $C[P] \Downarrow$ and the
thesis is proved.
\end{Proof}

\bigskip
\noindent
\textbf{\ccs\ is strongly replacement free.}\quad
To prove that \ccs\ is strongly rep-free (Theorem~\ref{teo:ccs-repfree}),
by Proposition~\ref{prop:simulate-rep-free}(1), it suffices to show
that $\osimul$ is a pre-congruence for \ccs.

\begin{proposition}
\label{theo:w-preserved}
Relation $\osimul$ is a pre-congruence for \ccs.
\end{proposition}
\begin{proofSketch}
First, for each operator of finite \ccs, it is proven by induction
on $k$ that $\simuln k$ is preserved for every $k$.
Then, it is proven that $\osimul$ is preserved by recursive definitions
by following the same strategy adopted in~\cite{CCS} for proving that
strong bisimulation is preserved by recursive definitions.
\end{proofSketch}

\begin{theorem}
\label{teo:ccs-repfree}
\ccs\ is strongly rep-free.
\end{theorem}
\begin{Proof}
Let $C$ be a context, $I$ be an invisible process, and $P$ be a process.
Since $I \osimul P$, by Proposition~\ref{theo:w-preserved}, we
get that $C[I]\osimul C[P]$. Then, the thesis follows by
Proposition~\ref{prop:simulate-rep-free}(1).
\end{Proof}

\bigskip
\noindent
\textbf{\pic\ is strongly replacement free.}\quad
To prove that \pic\ is a strongly rep-free calculus
(Theorem~\ref{theo:pcalc-repfree}), we proceed as for \ccs. However,
the $\omega$-simulation relation results to be a pre-congruence
w.r.t. all the operators of \pic\ but for the input prefix. This
happens because, intuitively, execution of an input action can
generate a substitution that identifies names that were originally
different. As a consequence of the application of this substitution
to the continuation process, new transitions could arise because
communications that were originally impossible become enabled, thus
the expansion of parallel composition in terms of choice and prefix
becomes unsound.

Another difference with the previous section is the proof technique
used. Indeed, to prove that $\osimul$ is preserved by all
operators but input prefix, we exploit the fact that our
$\omega$-simulation is strictly weaker than a well-known
observational semantics for \pic, named \emph{weak bisimilarity}
\cite{PICALC,PICALCBOOK} and denoted by $\approx$. This fact can be
easily proved by relying on a family of relations $\approx_k$
\cite[Def. 2.4.24, page 99]{PICALCBOOK}, that stratify the
definition of weak bisimilarity and whose intersection
$\approx_\omega$ is weaker than $\approx$
\cite[Theorem 2.4.27, page 100]{PICALCBOOK}
but stronger than $\osimul$.
Indeed, for every $k$, we have $\approx_k \subseteq \simuln{k}$
(it directly follows by definition since $\approx_k$ is
the greatest symmetric relation contained in $\simuln{k}$)
and this implies that $\approx_\omega \subseteq \osimul$.
Therefore,
for proving the following results about $\osimul$ we can exploit
all the laws that are sound for $\approx$.

Although in general input prefix does not preserve $\osimul$,
we can however prove, by induction on the structure of contexts,
that the premises of Proposition~\ref{prop:simulate-rep-free}(1) hold
(Lemma~\ref{cor-pic:w-preserved}) which suffices for
our purposes (Theorem~\ref{theo:pcalc-repfree}).

\begin{lemma}
\label{cor-pic:w-preserved}
For every context $C$, invisible process $I$ and process $P$,
it holds that $C[I] \osimul C[P]$.
\end{lemma}

\begin{theorem}
\label{theo:pcalc-repfree}
\pic\ is strongly rep-free.
\end{theorem}
\begin{Proof}
Let $C$ be a context, $I$ be an invisible process, and $P$ be a process. By
Lemma~\ref{cor-pic:w-preserved}, we have that $C[I] \osimul C[P]$.
Then, the thesis follows by Proposition~\ref{prop:simulate-rep-free}(1).
\end{Proof}

\section{Weakly replacement free calculi}
\label{sec:weakly-repfree}

In this section we deal with some calculi that are only weakly rep-free.
Intuitively, for the calculi we consider, violation of strong
replacement freeness is due to the introduction of one of three
different mechanisms: match among names, polyadic synchronization
and pattern matching. Among the many different process calculi
that have adopted these mechanisms,
for the sake of simplicity we analyze three known variants of \pic.
For each calculus we show that it is not strongly rep-free by
exhibiting a triple made of a context $C$, an invisible process $I$
and a process $P$ such that condition~(\ref{cond:rep-free}) of
Definition~\ref{def:rep-free} is violated.
In all the exhibited contexts the hole is in the scope of an input
prefix: thus, by means of an interaction, the context can generate a
substitution whose application to the enclosed invisible process
transforms it into a visible process by enabling a visible step
that was originally blocked.
Instead, if we restrain invisible processes only to those that are
closed, the strategy sketched above does not apply anymore.
To prove that the three calculi are weakly rep-free, instead of
three separate proofs, we define a richer variant, called \mpm,
incorporating them all and make the proof for it
(Theorem~\ref{theo:mpm-weakfree}).

\medskip
\noindent
\textbf{\pic\ with match.}\quad
The \emph{match} operator has been used in many
presentation of \pic, as \eg in the original one~\cite{PICALC}.
The following result trivially follows from Example~\ref{ex:picMatch}.

\begin{proposition}
\label{prop:pmatch-not-rep-free}
\pic\ with match operator is not strongly rep-free.
\end{proposition}

\medskip
\noindent
\textbf{\pic\ with polyadic synchronization.}\quad
This variant~\cite{Carbone-express-polyadic03} enriches \pic\ by
allowing to use \emph{tuples}, \ie sequences of names, denoted by
$\outDS{a_1, a_2, \ldots, a_n}$ or $\seq{a}$,
in addition to single names, for identifying input and output
channels. Thus, the syntax of input and output actions becomes
$\inpic{\seq a}{x}$ and $\outpic{\seq a}{n}$, respectively. The
operational semantics is defined by rules similar to those of \pic\
where subjects of input and output actions are tuples,
allowing interaction to happen only when such tuples match.

\begin{proposition}
\label{prop:poly-not-rep-free}
\pic\ with polyadic synchronization is not strongly rep-free.
\end{proposition}
\begin{Proof}
Let $C$, $I$ and $P$ be as follows:
$C = (\nu x)(\inpic{x}{a} . \indhole{} \mid \outpic{x}{b})$,
$I = (\nu z)(\outpic{\outDS{z, a}}{d} \mid \inpic{\outDS{z, b}}{w} . \outpic{y}{c})$
and $P = \nil$.
Process $I$ is invisible since synchronisation along channels
$\outDS{z, a}$ and $\outDS{z, b}$ cannot take place (as names $a$
and $b$ are supposed to be different). However, we have that $C[I]
\Downarrow$, in fact
\begin{center}
$
\begin{array}{r@{\ }c@{\ }l}
C[I] & = & (\nu x)(\inpic{x}{a} . (\nu z)(\outpic{\outDS{z, a}}{d}
\mid \inpic{\outDS{z, b}}{w} . \outpic{y}{c}) \mid \outpic{x}{b})
\transition \tau
(\nu x)((\nu z)(\outpic{\outDS{z, b}}{d} \mid \inpic{\outDS{z,
b}}{w} . \outpic{y}{c}) \mid \nil)
\transition \tau\\
& & (\nu x)((\nu z)(\nil \mid \outpic{y}{c}) \mid \nil)
\transition{\outpic{y}{c}}
(\nu x)((\nu z)(\nil \mid \nil) \mid \nil)
\end{array}
$
\end{center}
Instead, $C[P] \ghost$ since process $C[P]$ can only perform the
transition
$
C[P] = (\nu x)(\inpic{x}{a} \mid \outpic{x}{b}) \transition \tau
(\nu x)(\nil \mid \nil)
$
and become stuck in doing so.
\end{Proof}

\medskip
\noindent
\textbf{\pic\ with pattern matching.}\quad
This variant enriches \pic\ by allowing input actions to use
\emph{patterns of names} to selectively synchronise with output
actions along the same channel according to the offered tuples.
Pattern matching was introduced in the context of \pic\
in~\cite{Gorla-reasonable-main08},
but it is also used by several other process calculi,
like \eg those presented in~\cite{SOCK06,LPT07:ESOP}.
A pattern is a sequence of names where some names are placeholders
while some other ones are not and stand for themselves.
We indicate these latter ones by operator $\hood{\cdot}$.
Intuitively, a pattern $\seq x$ can match any tuple $\seq a$
having the same length obtained by instantiating the names
in $\fn{\seq x}$, where any name occurring in $\seq x$ is free but
for those that are argument of operator $\hood{\cdot}$, thus, e.g.,
$\hood x \sigma = \hood x$ for any name $x$ and substitution
$\sigma$. When the check if a pattern $\seq x$ and a tuple
$\seq a$ match succeeds, it returns the least substitution
$\sigma$ such that $\seq x \sigma = \seq a$ (once the occurrences of
operator $\hood{\cdot}$ in the left hand side have been removed).
Thus, a process $\inpic{a}{\seq x} . P$ and a process
$\outpic{a}{\seq b} . Q$ can synchronise if, and only if, ${\seq x}$
and ${\seq b}$ match by generating some substitution $\sigma$ and,
after the synchronisation, $\inpic{a}{\seq x} . P$ becomes $P
\sigma$. The operational semantics is defined by rules similar to
those of \pic\ where objects of input and output actions are
replaced by patterns and tuples, respectively, and the rule for
input prefix checks possible matching tuples.

\begin{proposition}
\label{prop:patmatch-not-rep-free}
\pic\ with pattern matching is not strongly rep-free.
\end{proposition}
\begin{Proof}
Let $C$, $I$ and $P$ be as follows:
$C = (\nu x)(\inpic{x}{a} . \indhole{} \mid \outpic{x}{b})$,
$I = (\nu z)(\outpic{z}{a} \mid \inpic{z}{\hood b} . \outpic{y}{c})$
and $P = \nil$.
Process $I$ is invisible since it is blocked because the pattern
$\hood b$ and the tuple $a$ do not match (as names $a$ and $b$
are supposed to be different).
However, we have that $C[I] \Downarrow$, in fact
\begin{center}
$
\begin{array}{r@{\ }c@{\ }l}
C[I] & = & (\nu x)(\inpic{x}{a} . (\nu z)(\outpic{z}{a} \mid
\inpic{z}{\hood b} . \outpic{y}{c}) \mid \outpic{x}{b})
\transition \tau
(\nu x)((\nu z)(\outpic{z}{b} \mid \inpic{z}{\hood b} .
\outpic{y}{c}) \mid \nil)
\transition \tau\\
& &
(\nu x)((\nu z)(\nil \mid \outpic{y}{c}) \mid \nil)
\transition{\outpic{y}{c}}
(\nu x)((\nu z)(\nil \mid \nil) \mid \nil)
\end{array}
$
\end{center}
Instead, $C[P] \ghost$ since process $C[P]$ can only perform the
transition
$
C[P] = (\nu x)(\inpic{x}{a} \mid \outpic{x}{b}) \transition \tau
(\nu x)(\nil \mid \nil)
$
and become stuck in doing so.
\end{Proof}

\medskip
\noindent
\textbf{\mpm.}\quad
As a consequence of Propositions~\ref{prop:pmatch-not-rep-free},
\ref{prop:poly-not-rep-free} and~\ref{prop:patmatch-not-rep-free},
we have that the three variants of \pic\ previously presented
are strictly more expressive than \pic, in other words
there is no basic encoding from any of them into \pic.
Now, we combine them to form a sort of super calculus,
that we call \mpm, corresponding to simultaneously adding
match, polyadic synchronization and pattern matching to \pic.
The syntax of \mpm\ is defined as
$$
\begin{array}{c}
P,Q \ \ ::= \ \
\nil \!\!\!\sep\!\!\! \mu . P \!\!\!\sep\!\!\! P + Q \!\!\!\sep\!\!\! P \!\mid\! Q \!\!\!\sep\!\!\! (\nu n) P \!\!\!\sep\!\!\! \reppic P \!\!\!\sep\!\!\! \mat{n}{m} P
\qquad\quad
\mu \ \ ::= \ \
\tau \!\!\!\sep\!\!\! \inpic{\seq a}{\seq x} \!\!\!\sep\!\!\! \outpic{\seq a}{\seq n}\\
\end{array}
$$
To save space, the presentation of the operational semantics
of \mpm\ is relegated to \cite{ExprFULL}.

We now show that \mpm\ is weakly rep-free from which
it follows that all the three variants of \pic\ we have
considered in this section are weakly rep-free too.

\begin{theorem}
\label{theo:mpm-weakfree}
\mpm\ is weakly rep-free.
\end{theorem}
\begin{proofSketch}
We prove that $C[I] \osimul C[P]$, for every context $C$, closed
invisible process $I$ and process $P$. The thesis then follows by
Proposition~\ref{prop:simulate-rep-free}(2).
Similarly to that of Lemma~\ref{cor-pic:w-preserved}, the proof is
an easy induction on the structure of context $C$ but replacing,
respectively, subjects and objects of input and output actions with
tuples and patterns, invisible processes with closed invisible
processes, and by exploiting the fact that if $I$ is a closed
invisible process, then $I\sigma$ is a closed invisible process too.
\end{proofSketch}

\section{Non replacement free calculi}
\label{sec:non-repfree}

In this section we show that most of the calculi with some form
of priority proposed in the literature, as \eg those presented
in~\cite{aceto-priority08,priority-guards08,priority1999,LPT07:ESOP},
are not rep-free. In process calculi, priority is one of the most
widely studied, and natural, notions used to implement different
levels of urgency between actions of (a system of) processes.
According to the terminology of \cite{priority1999} (that surveys
the different approaches taken in the literature), we consider
both calculi with local priority (as \eg
\cpg~\cite{priority-guards08}
and \cows~\cite{LPT07:ESOP}), and calculi with
global priority (as \eg \bccsp~\cite{aceto-priority08},
\ccsg\ and \ccsprio~\cite{priority1999}).

The results presented in this section demonstrate that there
exists no independence preserving basic encoding from any of
the calculi with priority into, e.g., \ccs\ or \pic, or
into any of the extensions of \pic\ we have presented in
Section~\ref{sec:weakly-repfree}.
For each calculus we show that it is not rep-free by
exhibiting a triple made of a context $C$, a closed invisible process $I$
(usually the null process $\nil$) and a process $P$ such that
condition~(\ref{cond:weak-rep-free}) of
Definition~\ref{def:weak-rep-free} is violated.

For the sake of simplicity, the fragments of the calculi
considered in this section will be slightly adapted and
simplified to avoid, as much as possible, introducing further
notations and complications.

\medskip
\noindent
\textbf{\bccsp.}\quad
\bccsp\ (BCCSP with the priority operator $\Theta$,~\cite{aceto-priority08})
is the simpler calculus with priority analyzed in this paper.
It is obtained by adding the well-known priority operator $\Theta$
of~\cite{Baeten:interrupt86} to the basic process algebra
BCCSP~\cite{Glabbeek:spectrum90}.
An utterly simplified syntax of \bccsp\ that, unlike the original
one~\cite{aceto-priority08}, does not allow action and process
variables, is as follows
\begin{center}
$P$ \ ::= \ $\nil \!\sep\! \mu . P \!\sep\! P + P  \!\sep\!
\Theta(P)$
\end{center}
The priority operator $\Theta$ gives certain actions priority over others
based on an irreflexive partial ordering relation $<$ over the set of actions.
Intuitively, $\mu < \mu'$ is interpreted as `$\mu'$ has priority over $\mu$'.
Thus, for example, if $P$ is some process that can initially perform both
$\mu$ and $\mu'$, then $\Theta(P)$ will not be able to initially
execute $\mu$. That is, in the context of the priority operator $\Theta$,
action $\mu$ is preempted by action $\mu'$.

\begin{proposition}
\label{prop:bccsp-not-rep-free}
\bccsp\ is not rep-free.
\end{proposition}
\begin{Proof}
Let $C$, $I$ and $P$ be as follows:
$C = \Theta(a + \indhole{})$,
$I = \nil$
and $P = \tau$,
with $a < \tau$. We have that $C[I] \Downarrow$, in fact $C[I] =
\Theta(a + \nil)$ can perform the transition
$
\Theta(a + \nil) \transition{a} \Theta(\nil)
$.
Instead, $C[P] \ghost$ since
$C[P] = \Theta(a + \tau)$ can
only perform the transition
$
\Theta(a + \tau) \transition{\tau} \Theta(\nil)
$
and become stuck in doing so.
\end{Proof}

It is worth noticing that even such an extremely simple calculus with
priority cannot be properly encoded into any rep-free calculus.

\medskip
\noindent
\textbf{\cpg.}\quad
\cpg\ (\ccs\ with Priority Guards,~\cite{priority-guards08}) is an
extension of \ccs\ allowing processes with \emph{priority guards},
\ie terms of the form $S:\mu.P$, where $S$ is some finite set
of visible actions,
which behave like $\mu.P$ except that action $\mu$ can only be
performed if the environment does not offer any action in
$\outccs{S}= \set{\outccs{n}\mid n\in S}$.
The syntax of the calculus is as follows
\begin{center}
$P$ \ ::= \ $\nil \!\sep\! \sum_{i \in I} S_i:\mu_i . P_i \!\sep\! P\ren{f} \!\sep\!
P\backslash L \!\sep\! P \mid Q \!\sep\! A\arr{a_1, \ldots, a_n}$
\end{center}
\cpg\ builds on a variant of \ccs\ where the choices are guarded
and parameterized process definitions are used in place of
recursion for modelling infinite behaviours.

A transition can be conditional on offers from the environment.
Consider $\set{a}:b \mid \outccs{b}$. It can make a computation
step due to the synchronization between $b$ and $\outccs{b}$.
However $b$ is guarded by $a$, and so the computation step
is conditional on the environment not offering $\outccs{a}$.
This is reflected by letting transitions be parameterised on labels
of the form $S:\mu$. The intended meaning of $P \transition{S:\mu}
P'$ is that $P$ can perform action $\mu$, and become $P'$ in doing
so, as long as the environment does not offer $\outccs{\alpha}$
for any $\alpha \in S$.

\begin{proposition}
\label{prop:CPG-not-rep-free}
\cpg\ is not rep-free.
\end{proposition}
\begin{Proof}
Let $C$, $I$ and $P$ be as follows:
$C = ((a + \set{a}:b . \outccs{c}) \mid\ \outccs{b} \mid \indhole{} )\backslash \set{a,b}$,
$I = \nil$
and $P = \outccs{a}$.
We have that $C[I] \Downarrow$,
in fact
$
C[I] \ = \ ((a + \set{a}:b . \outccs{c}) \mid\ \outccs{b} \mid \nil )\backslash \set{a,b}
\transition{\set{a}:\tau}
(\outccs{c} \mid\ \nil \mid \nil )\backslash \set{a,b}
\transition{\emptyset:\outccs{c}}
(\nil \mid\ \nil \mid \nil )\backslash \set{a,b}
$,
where $\emptyset:\outccs{c}$ is visible while $\set{a}:\tau$ is not.
Instead, $C[P] \ghost$ since process $C[P]$ can only perform the transition
$
C[P] = ((a + \set{a}:b . \outccs{c})\ \mid\ \outccs{b} \mid \outccs{a} )\backslash \set{a,b}
\transition{\emptyset:\tau}
(\nil \mid\ \outccs{b} \mid \nil )\backslash \set{a,b}
$
labelled by the invisible action $\emptyset:\tau$ and
become stuck in doing so.
\end{Proof}

The semantics of the fragment of \cpg\ where the only allowed
priority guarded processes are of the form $\emptyset:\mu.P$
coincides with that of \ccs.
Hence, \cpg\ is somehow more expressive than \ccs.
In~\cite{priority-guards08,Nadia-memorial-expr09} it is
also argued that expressiveness of \cpg\ and \pic\ is not
comparable.

\medskip
\noindent
\textbf{\ccsg\ and \ccsprio.}\quad
\ccsg\ (\ccs\ with static priority and global preemption,
\cite{priority1999}) is an extension of \ccs\ where channels have
priority levels and only complementary actions at the same level of
priority can engage in a communication. A notion of preemption then
stipulates that a process cannot engage in transitions labelled by
actions with a given priority whenever it is able to perform a
transition labelled by an internal action of a higher priority. In
this case, we say that the lower-priority transition is preempted
by the higher-priority internal transition. Therefore, visible
actions never have preemptive power over actions of lower priority
because visible actions only indicate the potential for execution.
For simplicity, we consider just two priority levels:
ordinary actions and higher priority, underlined actions.
The syntax of \ccsg\ is then the same as that of \ccs,
except for actions that can also be underlined.

When restricting only to unprioritized (or only to prioritized)
actions, the transition rules of \ccsg\ are exactly the ones of \ccs.
When both prioritized and unprioritized actions may be involved, an
unprioritized action is allowed only if no prioritized invisible
action can be executed.

\ccsprio~\cite{priority1999} extends \ccsg\ with two operators,
originally introduced in~\cite{HennessyCleaveland:IandC90}, which
correspond to the prioritization of a visible unprioritized action,
written $P\lceil \mu$, and to the deprioritisation of a visible
prioritized action, written $P\lfloor \underline\mu$.

\begin{proposition}
\label{prop:CCSsg-not-rep-free}
\ccsg\ and \ccsprio\
are not rep-free.
\end{proposition}
\begin{Proof}
The proof for \ccsg\ trivially follows from Example~\ref{ex:ccsg}.
To prove the statement for \ccsprio\ it is sufficient to take
$
C = (({\underline a} \mid \indhole{}) \backslash \set{\underline a} + \outccs{b}) \lceil b\,
$
and proceed similarly to the previous case.
\end{Proof}

Proposition~\ref{prop:CCSsg-not-rep-free} permits to conclude
that there exist no basic encodings of \ccsg\ into \ccs.
Since the fragment of \ccsg\ not containing prioritized actions
coincides with \ccs, and the identity encoding is a basic encoding
of \ccs\ into \ccsg, we obtain that \ccsg\ is strictly more
expressive than \ccs.

\medskip
\noindent
\textbf{\cows.}\quad
\cows\ (Calculus for Orchestration of Web Services,~\cite{LPT07:ESOP})
is a recent formalism specifically devised for modelling
service-oriented systems which integrates primitives of well-known
process calculi (\eg \pic) with constructs meant to model web services
orchestration (\eg communication endpoints and forced termination).
\cows\ is equipped with a priority mechanism that assigns
actions for forcing immediate
termination of concurrent processes greatest priority within their
enclosing scope. This way, when a fault arises in a scope, (some of)
the remaining processes of the enclosing scope can be terminated
before starting the execution of the relative fault handler.

Due to space limitations, we consider here only the very simple fragment
of the original calculus (without replication/choice/protection operators)
generated by the following syntax
\begin{center}
$P,Q$ \ ::= \
$\nil \!\sep\! \killing{\kappa} \!\sep\! \outpic{\seq a}{\seq n} \!\sep\! \inpic{\seq a}{\seq x} . P \!\sep\! \scope{\kappa} P \!\sep\! P \!\mid\! Q$
\end{center}
In addition to the set of names, we assume existence of a disjoint
set of \emph{killer labels}, ranged over by $\kappa$. They can be
used for introducing a named scope for grouping certain processes.
Being different from names,
killer labels cannot be exchanged in communications, thus their
scope is statically regulated by the \emph{delimitation}
operator $\scope{\kappa} P$.

We comment on the two novel operators, namely kill and
delimitation. $\killing{\kappa}$ causes immediate
termination of all concurrent processes inside an enclosing
$\scope{\kappa}$, that stops the killing effect by turning the
transition label $\kappa$ into $\tau$. Execution of parallel
processes is interleaved, but when a kill can be performed.
In fact, kill is executed \emph{eagerly} with respect to
the processes enclosed within the delimitation of the corresponding
killer label.

\begin{proposition}
\label{prop:COWS-not-rep-free} \cows\ is not rep-free.
\end{proposition}
\begin{Proof}
Let $C$, $I$ and $P$ be as follows:
$C = \scope{\kappa}(\, \indhole{} \  \mid \outpic{\seq a}{\seq n} \,)$,
$I = \nil$
and $P = \killing{\kappa}$.
We have that $C[I] \Downarrow$, in fact
$
C[I] = \scope{\kappa}(\, \nil \mid \outpic{\seq a}{\seq n} \,)
\transition{\outpic{\seq a}{\seq n}} \scope{\kappa}(\, \nil \mid
\nil \,)
$.
Instead, $C[P] \ghost$ since process $C[P]$ can only perform the
transition
$
C[P] = \scope{\kappa}(\, \killing{\kappa} \mid \outpic{\seq a}{\seq
n} \,) \transition \tau \scope{\kappa}(\, \nil \mid \nil \,)
$
and become stuck in doing so.
\end{Proof}

\section{Concluding remarks and related work}
\label{sec:conclusion}

To sum up, a first contribution of this paper is the introduction of
a metatheory based on the replacement freeness criterion and on the
notion of basic (possibly, independence preserving) encodings that
provides a method to tell process calculi apart.
Our approach is general and uniform enough to permit comparing the
relative expressive power of quite different process calculi.
On the contrary, many works on the subject only focus on variants
of the same calculus.
A second contribution consists in presenting a number of results coming
from the application of our metatheory to well-known process calculi
that are possibly extensions of \ccs\ or \pic.
We thus end up to retrieve separation results similar to, e.g.,
\cite{Gorla08:encodability-separation,Gorla-reasonable-main08}, but ours are stronger since they hold for a more general class of encodings,
or to, e.g.,
\cite{Carbone-express-polyadic03,priority-guards08,Nadia-memorial-expr09}, but here they follow by possibly simpler proofs.
Finally, some other results, e.g. some of those for non rep-free calculi presented in Section~\ref{sec:non-repfree}, as far as we know, are pointed out for the first time.

\smallskip
\noindent\textbf{Related work.\quad}
Defining an encoding of a process calculus into another one and
analysing the properties of the encoding function, or proving
that such an encoding cannot exist, is a widely adopted approach
for studying the relative expressive power of process calculi. This
is an effective approach that aims at comparing the calculi just on
the basis of their semantic differences, \ie without resorting to
any specific problem (\eg the `leader election' problem considered in \cite{EneM99,Palamidessi-expressive-synch03,PhillipsV06,VigliottiPP07,priority-guards08,PhillipsV08,Nadia-memorial-expr09}).
However, the significance of the obtained results is subordinated to
the properties that the encodings are required to enjoy, \ie to the
classes of encodings.

The most closely related works are \cite{Gorla-reasonable-main08},
that introduces the class of \emph{reasonable encodings} for
comparing several communication primitives in the context of \pic,
and \cite{Gorla08:encodability-separation}, that introduces the
class of \emph{valid encodings} for comparing many variants of \ccs,
\pic\ and Ambient calculus \cite{CardelliG00:ambient-calculus}.
Our basic encodings generalise the
reasonable encodings, the former ones being obtained from the latter
ones by dropping the conditions on name invariance, operational
correspondence and divergence preservation and reflection.
Similarly, our basic encodings generalise the valid encodings by
dropping the conditions on name invariance, operational
correspondence and divergence reflection. Indeed, \emph{success
sensitiveness} of \cite{Gorla08:encodability-separation} implies our
interaction sensitiveness requirement: appropriate observers can be
defined that test the capability of a process to interact with the
environment and report success only in that case. In fact,
\cite{Gorla-reasonable-main08,Gorla08:encodability-separation} aim
at identifying suitable criteria for both encodability and
separation results, thus the considered encodings are the outcome of
a compromise between `maximality' (typical of encodability results)
and `minimality' (typical of separation results),
while we are only interested to separation results and
the weaker the requirements on the encodings, the stronger
the results.
Moreover, we additionally consider process calculi with priority,
thus we end up establishing different separation results.

In fact, we impose somewhat coarser demands on our encodings than
those usually found in the literature.
For example, we don't require operational correspondence
(as instead done in \eg
\cite{Versari-bio-calculi07,Gorla-reasonable-main08,Gorla08:encodability-separation}),
observational correspondence or full abstraction (as instead done in
\eg \cite{GSV:fossacs04,DCE00,FuLu:10}), and in place of homomorphism (of \eg
parallel composition as required in
\cite{Carbone-express-polyadic03,Palamidessi-expressive-synch03,Versari-bio-calculi07,PhillipsV08})
we simply require compositionality. In particular, the requirement
of homomorphy for parallel composition 
(also called \emph{distribution preservation})
entails that the encoding
preserves the distribution of a term into parallel components
exactly, \ie without introducing any sort of coordinating context
that would reduce the degree of distribution. This requirement has
been sometimes criticized and indeed there exist encodings that do
not translate parallel composition homomorphically (see \eg
\cite{Nest00,BaldamusPV05}). Moreover, it is quite strong when
compared to the requirement of compositionality, imposed to ours
basic encodings, that only implies that any context in the source
calculus can be represented as a context in the target calculus.
Compositionality is a very natural property and, indeed, every
encoding we are aware of is defined compositionally.

Distribution preservation has been used as a requirement 
(for the encodings) in~\cite{priority-guards08,Nadia-memorial-expr09} 
for separating \cpg\ from both \ccs\ and \pic, thus obtaining results similar to our Proposition~\ref{prop:CPG-not-rep-free}.
It is a requirement for the class of \emph{uniform encodings} (that, other than distribution,
are required to preserve renaming, \ie to respect permutation of free names) introduced in~\cite{Palamidessi-expressive-synch03} for comparing the expressive power of synchronous and asynchronous versions of the \pic, which instead our criteria do not allow to separate. In fact, with a proof similar to, but simpler than, that of Theorem~\ref{theo:pcalc-repfree}, we can show that asynchronous \pic~\cite{ACS98} is rep-free as well.
An even stronger class of encodings
is used in~\cite{Carbone-express-polyadic03} to establish some separation results for variants of \pic. Among these results, the authors prove that match and polyadic synchronization cannot be encoded in \pic, which are similar to the separation results we present in Section~\ref{sec:weakly-repfree}.
Distribution preservation is also required for the encodings considered in~\cite{PhillipsV08}, where expressiveness of different variants of Ambient calculus is analysed, and in~\cite{Versari-bio-calculi07}, where an extension of the \pic\ with both polyadic synchronization and priority is introduced and used as target of `reasonable' encodings of a few
bio-inspired process calculi.

For some of the calculi with priority considered in this paper, there are already some separation results with respect to calculi without priority (see, e.g., \cite{Nadia-memorial-expr09,priority-guards08}). However, these results are given by considering either a less expressive fragment of \ccs\ as the target calculus or a strict class of encodings, typically uniform encodings.
In fact, distribution preservation seems even more restrictive when calculi with priority mechanisms
are involved. Indeed, priority alters the very basic notion of distributed computation, since in calculi with priority it is possible to know in advance if a process is not ready to perform some synchronisation \cite{Nadia-memorial-expr09}, which is instead not decidable in calculi without priority.
It seems then too demanding to require that the parallel operator of a calculus with priority exactly maps to the corresponding operator of a calculus without priority.

In~\cite{Aranda-Versari-expressivity09} the authors consider some
syntactic variants of \ccs\ with replication in place of recursion
and study the expressiveness of restriction and its interplay with
replication. They also enrich one of these variants with priority guards
\cite{priority-guards08} and conclude that priority adds
expressivity to this variant of the calculus. As comparison criteria
they consider decidability of convergence and relative
expressiveness with respect to a well-known observational semantics,
\ie \emph{failure semantics}.
Differently, we consider \ccs\ and its extension \cpg\ and
do not rely on any observational semantics.

\smallskip
\noindent\textbf{Future work.\quad}
For space limitation, in this paper we considered a subset of the calculi dealt with in \cite{ExprFULL}.
We intend to apply our metatheory to more process calculi, as \eg Psi-calculi~\cite{psi-calculi09}, that have already turn out to be quite expressive, and the prioritised variant of \pic\ introduced in~\cite{Versari-bio-calculi07}. Other more challenging applications, that might require an appropriate tuning of our metatheory, concern process calculi with communication mechanisms different from those considered in this paper, as \eg the broadcast variant of \pic\ considered in~\cite{VigliottiPP07}, Ambient calculus and higher order \pic\ \cite{PICALCBOOK}.
Moreover, we also plan to investigate the impact of loosening the
requirement of preserving name independence.
Indeed, although the requirement is used in many classes of encodings
(as \eg those used in \cite{Palamidessi-expressive-synch03,VigliottiPP07,priority-guards08,Nadia-memorial-expr09}), it leaves out of our study all those encodings that exploit some kind
of \emph{reserved} names for rendering specific primitives and
operators of the source calculus.
In this case, the encodings of independent processes might end up
not to be independent. Examples of such encodings can be found,
e.g., in \cite{Gorla-reasonable-main08,Gorla09:relative-expr}.
This may also lead to weakening the demand of compositionality for
allowing the encoding of a process of the source language to be
defined by combining the encodings of its subprocesses through a
single outermost context that coordinates their inter-relationships.

\smallskip
\noindent\textbf{Acknowledgements.\quad}
We thank Daniele Gorla and the anonymous reviewers for their fruitful comments that have helped us in improving the paper.

\bibliographystyle{eptcs}
\bibliography{biblio}

\end{document}